\documentclass[aps,prd,12pt,a4paper,groupedaddress,preprintnumbers,floatfix,nofootinbib,showpacs]{revtex4}
\usepackage[english]{babel}
\usepackage{amsmath}
\usepackage{graphicx}
\usepackage{color}
\usepackage{amssymb}
\usepackage{hyperref}
\usepackage{dcolumn}
\usepackage{bm}
\begin{document}
\preprint{DCP-11-04}

\title{Neutrino masses generation in a $Z_4$ model}

\author{
 Alfredo Aranda,$^{1,2}$\footnote{Electronic address:fefo@ucol.mx}
 Cesar Bonilla,$^{3}$\footnote{Electronic address:rasec.cmbd@gmail.com}
 and Alma D. Rojas$^{1}$\footnote{Electronic address:alma.drp@gmail.com}}

\affiliation{$^1$Facultad de Ciencias, CUICBAS,
Universidad de Colima, Colima, M\'exico \\
$^2$Dual C-P Institute of High Energy Physics, M\'exico\\
$^3$FCFM-BUAP, Puebla, M\'exico }

\date{\today}


\begin{abstract}
We present a renormalizable flavor model with $Z_4$ as flavor symmetry in both the quark and lepton sectors. 
The model is constructed with a minimal approach and no right-handed neutrinos are introduced. In this approach a minimum number of two SU(2) Higgs doublets  and one scalar singlet are required in order to obtain the Nearest Neighbor Interaction form for charged fermions and to generate neutrino masses radiatively.
  For the quark sector we follow the charge assignations made by Branco et al. in reference \cite{Branco:2010tx}.
All fermion masses and mixing angles in the model are in agreement with current experimental data and only the inverted hierarchy for the neutrino mass spectrum is allowed.
 Since neutrinos are Majorana the contribution to neutrinoless double beta decay is also analyzed.
\end{abstract}


\maketitle


\section{Introduction}
One of the main problems in the Standard Model (SM) is understanding the observed fermion masses and mixing angles.
 The fermion masses are determined by the flavor structure of Yukawa couplings after spontaneous symmetry breaking (SSB) but this is not restricted by the gauge symmetry. Another issue with the SM is the existence of neutrino masses as well as their nature (if they are Dirac or Majorana). These problems suggest we must go beyond the SM in an attempt to solve them.

In the quark sector we know the experimental values of the quark masses and the entries of the Cabibbo-Kobayashi-Maskawa (CKM) matrix \cite{Cabibbo:1963yz}.  It is not possible however to extract the flavor structure of quark mass matrices from the experimental values and the arbitrariness to choose the Yukawa matrix structures has lead model builders to study some particular textures, most specifically the so called Nearest Neighbour Interactions (NNI) texture \cite{Branco:2010tx,Harayama:1996am,NNI} and Fritzsch type texture \cite{Fritzsch:1977vd} (which assumes NNI structure together with hermiticity). The NNI structure assumes zeros in the (1,1), (1,3), (3,1) and (2,2) entries as shown below
\begin{equation}
\left(\begin{array}{ccc}
0 & * & 0\\
*&0&*\\
0&*&*
\end{array}\right).\label{NNI}
\end{equation}

Many groups have been used as flavor symmetries but discrete symmetries have been very successful describing  masses and mixing matrices. 
For instance, Branco et al. \cite{Branco:2010tx} showed that the NNI texture can be obtained, for the quark sector, through the introduction of an abelian flavor symmetry, and that the minimal realization, in the context of two SU(2) Higgs doublets, requires the introduction of the minimal symmetry $Z_4$. The remarkable aspect of this particular scenario is the fact that it works, including the lepton sector as we show below, with only two SU(2) doublet scalar fields, while for non-abelian symmetries, one usually requires a larger number of them and/or products of abelian and non-abelian groups (see~\cite{Aranda:2011dx} and references therein). Note that it was shown in \cite{Harayama:1996am}  that we can express the NNI matrix in terms of only four parameters, and that the Fritzsch type texture \cite{Fritzsch:1977vd,Fritzsch:1979zq} is a special case.

In the lepton sector the experimental measured values are the charged lepton masses, neutrino mass squared differences and mixing angles \cite{Nakamura:2010zzi}. The absolute neutrino mass scale is not known, nor their nature (if they are Majorana or Dirac type). The absolute neutrino mass squared differences and the mixing angles are determined from neutrino oscillation data, but these do not allow us to determine the sign of $\Delta m^2_{31(32)}$. In the case of 3 neutrinos mixing they could exhibit two possible mass spectrum: the so-called normal ($m_1<m_2<m_3$) and inverted hierarchies ($m_3<m_1<m_2$).

 The mixing matrix contains the three neutrino mixing angles and a CP violating phase\footnote{In the case of Dirac neutrinos we have only one CP violating phase, but if neutrinos are Majorana, there will be three CP violating phases.} is the mixing matrix $U_{PMNS}$. This matrix results from the product of the matrices that diagonalize charged leptons and neutrinos, and it can be parametrized in a similar way as the CKM matrix. 

In this sector other problems arise because we must obtain small neutrino masses and big mixing angles. Furthermore, the recent results on neutrino mixing angles obtained by the T2K experiment \cite{Abe:2011sj,Oyama:2011pt,Fogli:2011qn,Schwetz:2011zk} and  the Double Chooz experiment \cite{Abe:2011fz} have found evidence of a non-zero $\theta_{13}$ angle, contrary to the zero value assumed in some parametrizations, as is  the case of the tribimaximal parametrization \cite{Harrison:2002er}. This result  has increased the interest of flavor model builders whom have made some attempts to explain it \cite{T2K works}.

The most popular mechanism to generate neutrino masses is the see-saw mechanism \cite{Schechter:1980gr} but this requires the introduction of right handed neutrinos, a large number of scalar fields and/or energy scales. An alternative mechanism  that may or may not require right handed neutrinos is the radiative neutrino masses generation \cite{Zee:1980,Babu:1988qv,Fukugita:2003en}.

Renormalizable flavor models \cite{renormalizable} provide an option to avoid the large number of extra fields. In these models SU(2) scalar doublets, responsible of electroweak symmetry breaking, transform non-trivially under the flavor symmetry, and  could have interesting phenomenology at accessible energy scales.

 The alternative of generating neutrino masses radiatively has been recently explored in references \cite{Aranda:2011dx,Aranda:2010im},  where models are realized in the framework of renormalizable models with one flavor symmetry group and no right-handed neutrinos.
In those models the $Q_4$ group and the double tetrahedral group are considered as flavor symmetries, with a minimal approach, i. e., each model is constructed with the minimal SU(2) Higgs doublets and singlet scalar fields required to generate Majorana neutrino masses and the Fritzsch type texture for quarks and charged leptons. In both cases allowed mass matrices by experimental data are obtained and the contribution to neutrinoless double beta ($0\nu\beta\beta$) decay is also presented.

In general, the minimal number of Higgs doublets required  to generate neutrino masses radiatively at one loop is two. In reference \cite{Branco:2010tx} it was shown that it is possible to obtain the NNI form for the quark mass matrices, in the context of a two Higgs doublet extension of the SM, through the introduction of a $Z_4$ symmetry. In this scheme  they reproduce correctly the experimental allowed values  for quark masses and mixing angles.
Then, following the philosophy of references \cite{Aranda:2011dx,Aranda:2010im}, we take the charged lepton sector transforming under the flavor symmetry similarly as the down quark sector in  reference \cite{Branco:2010tx} and  we introduce the minimal extra-ingredients to generate  Majorana neutrinos masses radiatively, which agree with experimental data.

This work is organized as follows. In section II we introduce the model and generate neutrino masses. Then, in section III, a permutation symmetry in the lepton mass matrix is identified, which renders different textures with equivalent results. In section IV, the analysis to obtain mixing angles is performed and the contribution to the $0\nu\beta\beta$ decay is shown. Finally conclusions are presented.


\section{Fermion masses}

First, we present the model of Branco et al. \cite{Branco:2010tx} for quark sector. It requires two Higgs SU(2) doublets $\Phi_1$ and $\Phi_2$, with U(1) charges\footnote{This U(1) symmetry is imposed on the Lagrangian} $\phi_1\equiv Q(\Phi_1)$,    $\phi_2\equiv Q(\Phi_2)$. The charge assignments for quarks fields are:
\begin{equation}\label{quarks_charges}
 \begin{array}{rl}
  (q_1,q_2)=&(q_3+\phi_1-\phi_2,q_3-\phi_1+\phi_2),\\
(u_1,u_2,u_3)=&(q_3-\phi_1+2\phi_2,q_3+\phi_1,q_3+\phi_2),\\
(d_1,d_2,d_3)=&(q_3-2\phi_1+\phi_2,q_3-\phi_2,q_3-\phi_1),\\
 \end{array}
\end{equation}
where $q_i\equiv Q(Q_{Li})$, $u_i\equiv Q(u_{Ri}) $, $d_i\equiv Q(d_{Ri})$, and $Q_{Li}$ denoting the left handed quark doublets and $u_{Ri}$, $d_{Ri}$ the right handed quark singlets. The U(1) charges of the bilinear couplings $\bar{Q}_{Li}u_{Rj}$ and $\bar{Q}_{Li}d_{Rj}$ are, respectively,
\begin{equation}
\left(\begin{array}{ccc}
-2\phi_1+3\phi_2 & \phi_2  &-\phi_1+2\phi_2 \\
\phi_2& 2\phi_1-\phi_2 & \phi_1\\
-\phi_1+2 \phi_2 & \phi_1 & \phi_2
\end{array}\right),
\end{equation}
and
\begin{equation}
\left(\begin{array}{ccc}
-3\phi_1+2\phi_2 & -\phi_1  &-2\phi_1+\phi_2 \\
-\phi_1& \phi_1-2\phi_2 & -\phi_2\\
-2\phi_1+\phi_2 & -\phi_2 & -\phi_1
\end{array}\right).
\end{equation}
Given these expressions we see that promoting the U(1) to a $Z_4$ symmetry is required forbids some quark bilinear couplings and guarantees the zero textures for the NNI structure. Under this symmetry Higgs doublets must transform as
\begin{equation}
 \Phi_j\longrightarrow\Phi'_j=e^{\frac{\imath 2\pi}{4}\phi_j}\Phi_j,
\end{equation}
with
\begin{equation}
(\phi_1,\phi_2)=(1,2),
\end{equation}
and then, from relations in (\ref{quarks_charges}), the charge assignments for quarks are obtained :
\begin{eqnarray}\label{Qcharges}
 (q_1,q_2,q_3)=(2,0,3),\\
(u_1,u_2,u_3)=(2,0,1),\\
(d_1,d_2,d_3)=(3,1,2).
\end{eqnarray}
With these ingredients, the most general Yukawa couplings allowed by the $Z_4$ symmetry are given by
\begin{equation}\label{L_quarks}
-\mathcal{L}_{q}=\Gamma^1_u \bar{Q}_L\tilde{\Phi}_1 u_R+\Gamma^2_u \bar{Q}_L\tilde{\Phi}_2 u_R+\Gamma^1_d \bar{Q}_L{\Phi}_1 d_R+\Gamma^2_d \bar{Q}_L {\Phi}_2 d_R+h.c.,
\end{equation}
where $\Gamma^{1,2}_{u,d}$ are the Yukawa matrices:
\begin{equation}\label{quarkGammas}
\Gamma^1_u=\left(\begin{array}{ccc}
0 &0 & 0\\
0&0&b_u\\
0&b'_u&0
\end{array}\right),\qquad\Gamma^2_u=\left(\begin{array}{ccc}
0 & a_u & 0\\
a'_u&0&0\\
0&0&c_u
\end{array}\right),
\end{equation}
\begin{equation}\label{quarkGammas1}
\Gamma^1_d=\left(\begin{array}{ccc}
0 & a_d & 0\\
a'_d&0&0\\
0&0&c_u
\end{array}\right),\qquad
\Gamma^2_d=\left(\begin{array}{ccc}
0 &0 & 0\\
0&0&b_d\\
0&b'_d&0
\end{array}\right).
\end{equation}

When the Higgs doublets acquire their vacuum expectation values  $\left\langle \Phi_1 \right\rangle =v_1  $,  $\left\langle \Phi_2 \right\rangle=v_2 $, the NNI mass matrices are generated
\begin{equation}\label{mqu}
M_u=\left(\begin{array}{ccc}
0 & a_u v_2 & 0\\
a'_uv_2&0&b_uv_{1}\\
0&b'_uv_{1}&c_uv_{2}
\end{array}\right),
\end{equation}
\begin{equation}\label{mqd}
M_d=\left(\begin{array}{ccc}
0 & a_d v_1 & 0\\
a'_dv_1&0&b_dv_{2}\\
0&b'_dv_{2}&c_d v_{1}
\end{array}\right).
\end{equation}

Now, in the lepton sector we choose the same charge assignments of the down type quarks for the lepton SU(2) doublets ($L_{Li}$) and singlets ($l_{Ri}$):
\begin{eqnarray}\label{Lcharges}
 (\alpha_1,\alpha_2,\alpha_3)=(2,0,3),\\
\label{lcharges}
(e_1,e_2,e_3)=(3,1,2),
\end{eqnarray}
where  $\alpha_i\equiv Q(L_{Li}) $ and $e_i\equiv Q(l_{Ri})$.
With these charges, the general Yukawa couplings allowed by the $Z_4$ symmetry are:
\begin{equation}\label{L_leptons}
-\mathcal{L}_{leptons}=(\Gamma^1_l)_{ij} \bar{L_{Li}}\Phi_1 l_{Rj}+(\Gamma^2_l)_{ij} \bar{L_{Li}}\Phi_2 l_{Rj}+h.c.,
\end{equation}
where we have omitted SU(2) indices, $i,j$ are family indices and  $\Gamma_l^1,\Gamma_l^2$ are given by
\begin{equation}
\Gamma^1_l=\left(\begin{array}{ccc}
0 & a_l & 0\\
a'_l&0&0\\
0&0&c_l
\end{array}\right),\qquad
\Gamma^2_l=\left(\begin{array}{ccc}
0 &0 & 0\\
0&0&b_l\\
0&b'_l&0
\end{array}\right).
\label{Gammas}
\end{equation}
After spontaneous gauge symmetry breaking  we are left with the mass matrix
\begin{equation}\label{ml}
M_l=\left(\begin{array}{ccc}
0 & a_l v_1 & 0\\
a'_lv_1&0&b_lv_{2}\\
0&b'_lv_{2}&c_lv_{1}
\end{array}\right).
\end{equation}

Now, in order to generate neutrino masses radiatively \cite{Zee:1980,Fukugita:2003en,Aranda:2011dx,Aranda:2010im} we need to introduce some scalar singlets under the SM, $h$, with hypercharge $y=-1$, lepton number $L=2$, and appropiate charges under $Z_4$. 
 In order to identify the $Z_4$ charges of these fields we observe that the Yukawa couplings of $h$ must be of the form
\begin{equation}
 -\mathcal{L}_{LLh}=\kappa^{ab} \epsilon_{ij} \overline{(L^a_{Li})^c}L^b_{Lj} h^{*} +h.c.,
\end{equation}
where $i,j,$ are SU(2) indices, $\epsilon_{ij} $ is the antisymmetric tensor, $a,b$ are family indices and $\kappa$ is antisymmetric by Fermi statistic, $\kappa^{ab}=-\kappa^{ba}$ . Also, in the scalar sector there will be cubic coupling terms of this scalar singlet with the Higgs doublets
\begin{equation}
 -\mathcal{L}_{\Phi\Phi h}=\lambda_{\alpha\beta} \epsilon_{ij} \Phi_i^{\alpha}\Phi_j^{\beta} h +h.c.,
\end{equation}
where $\lambda_{\alpha\beta}=-\lambda_{\beta\alpha}$, $\alpha,\beta=1,2$.

With the charge assignments in (\ref{Lcharges}), the charges of the bilinear couplings  $\overline{(L^a_{L})^c}L^b_{L}$ under U(1) are 
\begin{equation}
\left(\begin{array}{ccc}
0 & 2 & 1\\
2&0&3\\
1&3&2
\end{array}\right),\label{LLcharges}
\end{equation}
while for the bilinear couplings $\Phi^{\alpha}\Phi^{\beta}$ we obtain
\begin{equation}
\left(\begin{array}{cc}
2 & 3 \\
3&0
\end{array}\right).\label{PhiPhicharges}
\end{equation}
  
\begin{figure}[ht]
	\centering
	\includegraphics[width=0.6\textwidth]{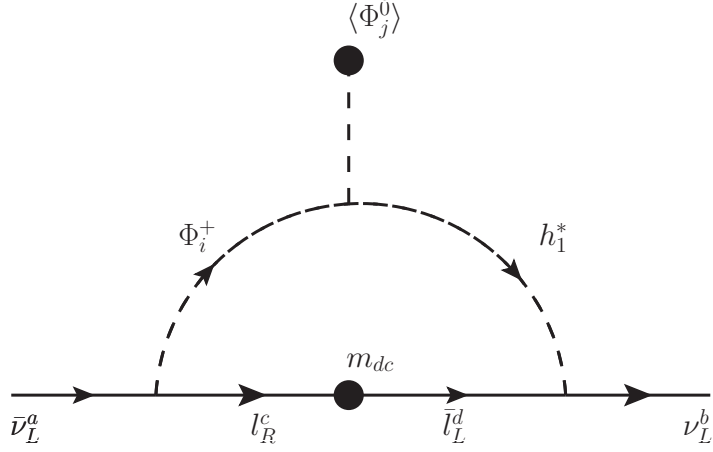}
	\caption{Diagram for radiative corrections to neutrino masses ($i,j=1,2, i\neq j$, and $m_{dc}$ are the matrix elements of $M_l$).}
	\label{loop}
\end{figure}

Then, from the allowed options by the $Z_4$ symmetry\footnote{In principle, we could have three scalar singlets $h_i$, $i=1,2,3$, with charges combination of $Q(h_1)=+1,-3$, $Q(h_2)=+2,-2$ and $Q(h_3)=+3,-1$, but the only one that provides couplings with $\Phi_i\Phi_j$ is $h_1$. }, 
we are left with only one scalar singlet, $h_1$, with charge $Q(h_1)=1$ or $Q(h_1)=-3$ (they are equivalent mod(4)). This is the only option which has the necessary couplings to construct the loop diagram shown in Figure \ref{loop}. With any of the two possible charges for $h_1$ we obtain the following couplings terms
\begin{equation}
 -\mathcal{L}_{LLh}=\kappa^{ab}_1 \epsilon_{ij} \overline{(L^a_{iL})^c}L^b_{Lj} h_1^{*} +h.c.,\label{L_LLh}
\end{equation}
where 
\begin{equation}
\kappa_1=\left(\begin{array}{ccc}
0 & 0 &\kappa\\
0&0&0\\
-\kappa&0&0
\end{array}\right),\label{kappa}
\end{equation}
and
\begin{eqnarray}
 -\mathcal{L}_{\Phi\Phi h}&=&\lambda_{\alpha\beta} \epsilon_{ij} \Phi_i^{\alpha}\Phi_j^{\beta} h_1 +h.c., \label{L_PhiPhih}\\
&=&2\lambda_{12}\Phi_1^{+}\Phi_2^0 h_1 - 2\lambda_{21}\Phi_2^{+}\Phi_1^0 h_1 +h.c.
\end{eqnarray}

Using these elements, together with the Yukawa coefficients matrix
\begin{equation}
Y_l=\left(\begin{array}{ccc}
0&a_l&0\\
a'_l&0&b_l\\
0&b'_l&c_l
\end{array}\right),\label{yl}
\end{equation}
that we obtain from expressions (\ref{L_leptons}), (\ref{Gammas}) and (\ref{ml}), we can construct loop interactions as those shown on Figure \ref{loop}.

First, we diagonalize matrix $M_l$  in expression (\ref{ml}) as\footnote{In order to perform the diagonalization we base on reference \cite{Harayama:1996am} and  set $a_l=a'_l$}
\begin{equation}
U_L M_lU_R^\dag=\text{diag}(m_e,m_\mu,m_\tau)
\label{mldiag}
\end{equation}
where $U_L$ can be obtained by diagonalizing $M_l^2=M_lM_l^\dag$ as
\begin{equation}
U_L M_l^2 U_L^\dag=\text{diag}(m_e^2,m_\mu^2,m_\tau^2)\label{ml2diag},
\end{equation}
and then used to solve for $U_R$ in equation (\ref{mldiag}). We can write a matrix for the lepton line on Figure \ref{loop} on the charged lepton diagonal mass basis as
\begin{equation}\label{uymku}
U_L Y_l U_R U_R^\dag M_l^\dag U_L U_L^\dag \kappa U_L^\dag=U_LY_l  M_l^\dag \kappa U_L^\dag .
\end{equation}
Then, we can obtain the interactions involved in the diagram loop by examining the trilinear scalar potential and the matrix  $Y_l  M_l^\dag \kappa$. To obtain the total contribution for each non-diagonal entry of $M_{\nu}$, given its symmetry, we need to add also the contribution of the transpose loop diagram. The matrix $M_{\nu}$ obtained in this way is rotated with $U_L$ to obtain the neutrino mass matrix in the charged lepton diagonal mass basis, $M'_{\nu}$. 

 The resulting entries in the Majorana neutrino mass matrix $M_{\nu}$ are (before rotating with $U_L$, i.e., not yet in the charged lepton diagonal mass basis)  
\begin{align}
m_{\nu_e\nu_e}=&2a_l\kappa^{31}m_{\tau\mu}\lambda_{12}v_2F(m_{\Phi}^2,m_h^2)\label{nu_11},\\
m_{\nu_\mu\nu_\mu}=&0\label{nu_22},\\
m_{\nu_\tau\nu_\tau}=&2b'_l\kappa^{13}m_{e\mu}\lambda_{21}v_1F(m_{\Phi}^2,m_h^2)\label{nu_33},\\
m_{\nu_e\nu_\mu}=&m_{\nu_\mu\nu_e}= 2b_l\kappa^{31}m_{\tau\tau}\lambda_{21}v_1F(m_{\Phi}^2,m_h^2)\label{nu_21},\\
m_{\nu_e\nu_\tau}=&m_{\nu_\tau\nu_e}=2(a_l\kappa^{13}m_{e\mu}\lambda_{12}v_2+b'_l\kappa^{31}m_{\tau\mu}\lambda_{21}v_1+c_l\kappa^{13}m_{\tau\tau}\lambda_{12}v_2) F(m_{\Phi}^2,m_h^2)\label{nu_13},\\
m_{\nu_\mu\nu_\tau}=&m_{\nu_\tau\nu_\mu}=0\label{nu_23},
\end{align}
where $ F(m_{\Phi}^2,m_h^2)$  is a scalar loop factor given by
\begin{equation}
  F(m_{\Phi}^2,m_h^2)=\frac{1}{16\pi^2}\frac{1}{m_{\Phi}^2-m_h^2}\log \frac{m_{\Phi}^2}{m_h^2},
\end{equation}
with $m_{\Phi}^2$  and $m_h^2$ denoting the charged Higgs and singlet field $h$ masses.
This gives the texture
\begin{equation}
\label{mnu}
M_\nu= \left(\begin{array}{ccc}
A&B&C\\
B&0&0\\
C&0&D\\
\end{array}\right).
\end{equation}

We must say here that in a recent work \cite{Fritzsch:2011qv} Fritzsch et al. excluded a two zero texture\footnote{It is a two zero texture because of the symmetry of the Majorana neutrino mass matrix.} like this, but there is not any inconsistency because the neutrino mass matrix that they refer is already in the basis where charged lepton mass matrix is diagonal, and at this point our neutrino mass matrix, $M_{\nu}$, is still in the weak basis where charged leptons have the NNI form.


\section{Other textures}

Considering  $a_l=a'_l$  in our NNI matrix form\footnote{This is justified by reference \cite{Harayama:1996am} where they show that the NNI form can be parametrized by four independent parameters only.} we obtain a Fritzsch-like texture for the lepton mass matrix
\begin{equation}
M_l=\left(\begin{array}{ccc}
0&a_l&0\\
a_l&0&b_l\\
0&d_l&c_l
\end{array}\right),\label{Ml}
\end{equation}
then, 
\begin{equation}
M_l^2\equiv M_lM_l^{\dag}=\left(\begin{array}{ccc}
a^2_l&0&a_ld_l\\
0&a_l^2+b_l^2&b_lc_l\\
a_ld_l&b_lc_l&d_l^2+c_l^2
\end{array}\right),\label{Ml2}
\end{equation}
where $a_l$, $b_l$, $c_l$, $d_l$ are real parameters (we have assumed the phases to be zero). We observe that any permutation of columns in matrix (\ref{Ml}) will give us the same  matrix (\ref{Ml2}). The matrices with this property are:
\begin{equation}
 \begin{array}{ccccc}
  \left(\begin{array}{ccc}
0&0&a_l\\
a_l&b_l&0\\
0&c_l&d_l
\end{array}\right)
 &\left(\begin{array}{ccc}
a_l&0&0\\
0&a_l&b_l\\
d_l&0&c_l
\end{array}\right)&\left(\begin{array}{ccc}
a_l&0&0\\
0&b_l&a_l\\
d_l&c_l&0
\end{array}\right)&\left(\begin{array}{ccc}
0&0&a_l\\
b_l&a_l&0\\
c_l&0&d_l
\end{array}\right)&\left(\begin{array}{ccc}
0&a_l&0\\
b_l&0&a_l\\
c_l&d_l&0
\end{array}\right).
\end{array}\label{permutations}
\end{equation}

In order to obtain each matrix we would need to do a new charge assignment for leptons, permuting also the charge assignment that we made for singlets $l_{Ri}$ only. We had
\begin{eqnarray}
Q(L_{Li}):\quad (\alpha_1,\alpha_2,\alpha_3)=(2,0,3),\\
Q(l_{Ri}):\quad (e_1,e_2,e_3)=(3,1,2),
\end{eqnarray}
with $Q(\Phi_i):$ $(\phi_1,\phi_2)=(1,2)$.
 Then, for instance, to get the first matrix in (\ref{permutations}) we rearrange the charges in this way: $(e_1,e_2,e_3)=(3,2,1)$. In this case, the charges for bilinear couplings are
\begin{equation}
\left(\begin{array}{ccc}
1&0&-1\\
3&2&1\\
0&-1&-2
\end{array}\right).
\end{equation}

The Yukawa couplings allowed by the $Z_4$ symmetry have the same form than in equation (\ref{L_leptons})
\begin{equation}
 -\mathcal{L}_{leptons}=\Gamma^1_{lA} \bar{L_L}\Phi_1 l_R+\Gamma^2_{lA} \bar{L_L}\Phi_2 l_R+h.c.,\nonumber
\end{equation}
but now
\begin{equation}
\Gamma^1_{lA}=\left(\begin{array}{ccc}
0 & 0 & a_l\\
a'_l&0&0\\
0&c_l&0
\end{array}\right),\qquad
\Gamma^2_{lA}=\left(\begin{array}{ccc}
0 &0 & 0\\
0&b_l&0\\
0&0&b'_l
\end{array}\right).
\end{equation}

To generate neutrino masses, as the lepton doublets charges remain unchanged, we introduce again only one scalar singlet $h_1$ with $L=2$, $Y=-1$ and $Q(h_1)=1$.

For couplings $\bar{(L)^c}L h$ and $\Phi\Phi h$ we obtain the same Lagrangians as in equations (\ref{L_LLh}) and (\ref{L_PhiPhih}) respectively. 
And then, the result for neutrino masses is essentially the same, but replacing indices $\mu\leftrightarrow\tau$  in equations (\ref{nu_11}-\ref{nu_23}).

In any case, the most general renormalizable scalar potential $V$, compatible with $Z_4$ symmetry and gauge symmetry,  is the same as in reference \cite{Branco:2010tx}, $V_1$, plus extra terms including the scalar singlet $h_1$
\begin{equation}
 V=V_1+V_2+V_3,
\end{equation}
where
\begin{eqnarray}
 V_1&=&\mu_1|\Phi_1|^2+\mu_2|\Phi_2|^2+\lambda_1|\Phi_1|^2+\lambda_2|\Phi_2|^2\\
&&+\lambda_3|\Phi_1|^2|\Phi_2|^2+\lambda_4\Phi_1^{\dag}\Phi_2\Phi_2^{\dag}\Phi_1,
\end{eqnarray}
and
 \begin{eqnarray}
  V_2&=&\mu_3 |h_1|^2+\lambda_5 |h_1|^2|\Phi_1|^2+\lambda_6 |h_1|^2|\Phi_2|^2,\\
V_3&=&\lambda_{\alpha\beta} \epsilon_{ij} \Phi_i^{\alpha}\Phi_j^{\beta} h_1 +h.c.
 \end{eqnarray}
Here we have included the trilinear coupling given in expression (\ref{L_PhiPhih}). As sugested in \cite{Branco:2010tx}, the inclusion of $h_1$ avoids the global symmetry acquired accidentally by $V_1$, in particular,  the terms in $V_3$ play the alternative role to the soft-breaking term of the $Z_4 $ symmetry that they introduce.


\section{Mixing angles for the lepton sector}
To make the analysis we rewrite the matrix $M_l$ in (\ref{ml}) as
\begin{equation}\label{M_l_2}
 \left(
\begin{array}{ccc}
 0 &A_l & 0 \\
 A_l & 0 & B_l \\
 0 &D_l &y_l^2 m_{\tau}
\end{array}
\right),
\end{equation}
where $A_l$, $B_l$,  $D_l$ and $y_l$ are real parameters, and $ m_{\tau}$ is the $\tau$ lepton mass.

Diagonalizing $M_l^2$ and solving for the three charged lepton masses one finds
\begin{equation}
  U_L=\left(
\begin{array}{ccc}
 0.997544 & 0.0672274 & 0.0196681 \\
 0.0700398 & -0.960886 & -0.267942 \\
 -0.000885754 & -0.268661 & 0.963234
\end{array}
\right),
\end{equation}
with the remaining free parameter chosen to be
 $y_l=0.969$ (note that there is a range for $y_l$ where the fit works, namely $0.0696 \leq |y_l| \leq 0.969$), and for the matrix $U_{R}$
\begin{equation}
U_R=\left(
\begin{array}{ccc}
 -0.997619 & -0.0672224 & 0.0154319 \\
 0.0689614 & -0.975912 & 0.206979 \\
 0.00114655 & 0.20755 & 0.978224
\end{array}
\right).
\end{equation}

To find the mixing in the lepton sector, $M_\nu$ is rotated with $U_L$
\begin{equation}
 M_\nu '= U_L M_\nu U_L^\dag,
\end{equation}
obtaining in this form the neutrino mass matrix in the basis where the charged lepton mass matrix is diagonal,  $M_\nu'$.    $M_\nu$ is the Majorana neutrino mass matrix that we found has the texture
\begin{equation}
M_\nu= \left(\begin{array}{ccc}
A&B&C\\
B&0&0\\
C&0&D\\
\end{array}\right).
\end{equation}

 The neutrino mixing is then obtained by diagonalizing  $M_\nu'$
\begin{equation}
 M_\nu'=V M_{\nu}^{Diag}V^{T},
\end{equation}
with $M_{\nu}^{Diag}$ representing the diagonal neutrino mass matrix and  $V$ the flavor mixing matrix. 

Because in this model neutrinos are Majorana fermions, it is convenient  to express $V$ \cite{BenTov:2011tj,Fritzsch:2011qv} as the product $V=U_{PMNS}P$, where $U_{PMNS}$ is the $3\times 3 $ unitary matrix containing the three flavor mixing angles $(\theta_{12},\theta_{23},\theta_{13})$ and one  CP-violating phase $\delta_{CP}$, and $P$ is a diagonal matrix containing two Majorana CP-violating phases $(\sigma,\rho)$, $P\equiv diag(e^{\imath\sigma},e^{\imath\rho},1)$. We adopt the parametrization
\begin{equation}
\label{Umns}
U_{PMNS}= \left(\begin{array}{ccc}
c_{13}c_{12}&c_{13}s_{12}&s_{13}\\
-c_{23}s_{12}e^{-\imath\delta_{CP}}-s_{23}s_{13}c_{12}&c_{23}c_{12}e^{-\imath\delta_{CP}}-s_{23}s_{13}s_{12}&s_{23}c_{13}\\
s_{23}s_{12}e^{-\imath\delta_{CP}}-c_{23}s_{13}c_{12}&-s_{23}c_{12}e^{-\imath\delta_{CP}}-c_{23}s_{13}s_{12}&c_{23}c_{13}\\
\end{array}\right),
\end{equation}
where $c_{ij}\equiv\cos \theta_{ij}$, $s_{ij}\equiv\sin\theta_{ij}$.  
We assume  that CP is conserved in this sector, thus $\delta_{CP}=0 $.
With this parametrization the neutrino mass matrix can equivalently be written as
\begin{equation}
\label{Mnulambda}
M'_\nu= U_{PMNS}\left(\begin{array}{ccc}
\lambda_1&0&0\\
0&\lambda_2&0\\
0&0&\lambda_3\\
\end{array}\right)U_{PMNS}^{T},
\end{equation}
where $\lambda_1=m_1e^{2\imath\sigma}$, $\lambda_2=m_ 2e^{2\imath\rho}$ and $\lambda_3=m_3$, being $m_i$, $i=1,2,3$ the positive real neutrino masses.

At this point it is convenient to note that we are assuming Majorana mass matrix elements of $\mathcal{O}$(
eV). This can be obtained, for example, observing that in equations (\ref{nu_11}-\ref{nu_23}) the parameters $a_l v_2,b_l v_1, b'_l v_1,c_lv_2 $ must be at the same scale of the lepton masses $m_l$. Then, if we assume $\lambda_{12} \sim m_{\Phi}\sim 500$ GeV and $\kappa\sim \mathcal{O}$(1), with $m_h\sim 4\times 10^{5}$GeV, this yields to mass matrix entries of  $\mathcal{O}$(eV). 

To perform the numerical analysis we used the results from the last global neutrino data analysis \cite{Schwetz:2011zk} 
\begin{equation}
\begin{array}{lll}
\sin^2 \theta_{12}=& 0.312^{+0.017}_{-0.015},&\\
\sin^2\theta_{23}=& 0.52^{+0.06}_{-0.07}&(0.52\pm 0.06),\\
\sin^2\theta_{13}=& 0.013^{+0.007}_{-0.005}&(0.016^{+0.008}_{-0.006}),\\
 \end{array}
\end{equation}
with  $\delta_{CP}=0$ and normal (inverted) hierarchy.

For the mass squared difference of neutrino masses we used also the parameters from the global fit \cite{Schwetz:2011zk}  
\begin{eqnarray}\label{Deltam21}
 \Delta m^2_{21}=&7.59^{+0.020}_{-0.18}\times 10^{-5} \text{eV}^2,&\\
\label{Deltam32}
 \Delta m^2_{32}=&2.50^{+0.09}_{-0.16}\times 10^{-3} \text{eV}^2 &(-2.40^{+0.08}_{-0.09}
\times 10^{-3} \text{eV}^2).
\end{eqnarray}

Given our lack of information about the absolute mass scale or neutrinos,
we used the following range for the square mass ratio
\begin{equation}\label{ratio}
\text{NH (IH):  } 0.029(0.030) < \left |\frac{\Delta m^2_{21}}{\Delta m^2_{32}}\right| < 0.032(0.033),
\end{equation}
obtained summing in quadrature the relative errors of $\Delta m^2_{21}$ and $|\Delta m^2_{32}|$.
   
In order to determine if the mass matrices reproduce the allowed experimental values for mass ratios and mixing angles, we perform a scan over the complete range of all three angles.

\begin{figure}[h]
	\centering
		\includegraphics[width=0.5\textwidth]{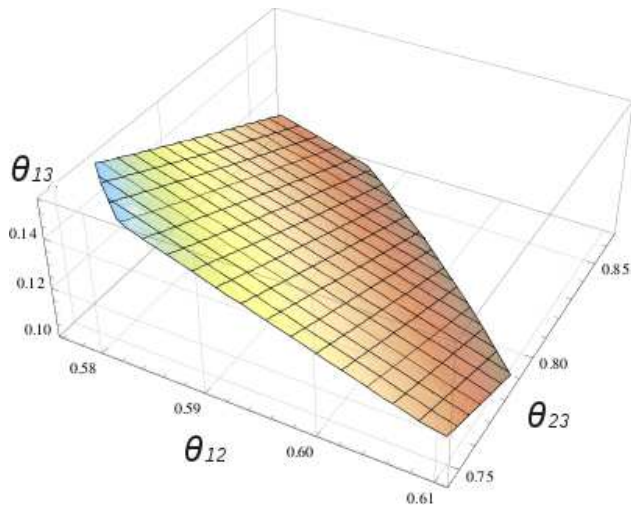}
		\caption{Angles that repeat the experimental mass difference ratio for the neutrino sector (Inverted hierarchy).}
      \label{lepangNeg}
\end{figure}

The angles obtained from our model that give (simultaneously) a ratio that falls within its allowed experimental range are shown on Figure \ref{lepangNeg}

From the analysis we can conclude that:
\begin{enumerate}
\item We only can have inverted hierarchy (IH), $m_3<m_2$~\footnote{As we are using the ratio (\ref{ratio}) to select the angles provided by our model from the allowed experimental ranges, the ratio could be positive or negative, but those with positive ratio are no compatible with the well established relation (\ref{Deltam21}).}.
From the angles that satisfy the mass ratio (\ref{ratio}) and from the requirement of $m_3>0$ in equation (\ref{Mnulambda}), our texture give us the values $\sigma  =\pi/2$ and $\rho =0$ for Majorana CP phases. We obtained these by replacing the angles in the diagonalization condition (\ref{Mnulambda}) and solving for the parameters $A,B,C,D$ in matrix $M_{\nu}$ in (\ref{mnu}).

\item The angles, for IH, have the bounds
\begin{equation}
 \begin{array}{c}
0.297<\sin^2(\theta_{12})<0.329,\\
0.46<\sin^2(\theta_{23})<0.58,\\
0.01<\sin^2(\theta_{13})<0.024.\\
 \end{array}
\end{equation}

\item We can see that the angles have solutions for all the experimental range. The values on $\theta_{13}$ are non zero and they agree with the recent results favoured by the T2K experiment\cite{Abe:2011sj}.
\end{enumerate}

Furthermore, any model beyond the standard model, which allow for lepton number violation, potentially contributes to $0\nu\beta\beta$ decay, and since in our model neutrinos are Majorana, $0\nu\beta\beta$ can take place. The amplitude of this decay is proportional to $|(M'_{\nu})_{ (11)}|$, i.e. the element (1,1) of the neutrino mass matrix in the charged lepton diagonal mass basis, which can be written as \cite{BenTov:2011tj,Hirsch:2006tt}
\begin{equation}
( M'_{\nu})_{ (11)}\equiv m_{\beta\beta}=e^{2\imath\sigma}\cos^2 \theta_{12}\cos^2 \theta_{13}m_1 +e^{2\imath\rho}\sin^2 \theta_{12}\cos^2 \theta_{13}m_2+\sin^2 \theta_{13}m_3,
\end{equation}
where $m_j$, $j=1,2,3$, are the real masses of neutrinos and $\sigma,\rho$ the Majorana phases. For the case of IH, $m_3\leq m_1\leq m_2$, we can rewrite this as
\begin{equation}
\begin{array}{ccl}
   m_{\beta\beta}&=&e^{2\imath\sigma}\cos^2 \theta_{12}\cos^2 \theta_{13}\sqrt{m_3^2+|\Delta m^2_{32}|-\Delta m^2_{21}} +e^{2\imath\rho}\sin^2 \theta_{12}\cos^2 \theta_{13}\sqrt{m_3^2+|\Delta m^2_{32}}| \\
&&+\sin^2 \theta_{13}m_3.
\end{array}
\end{equation}

\begin{figure}
	\centering
		  \includegraphics[width=0.7\textwidth]{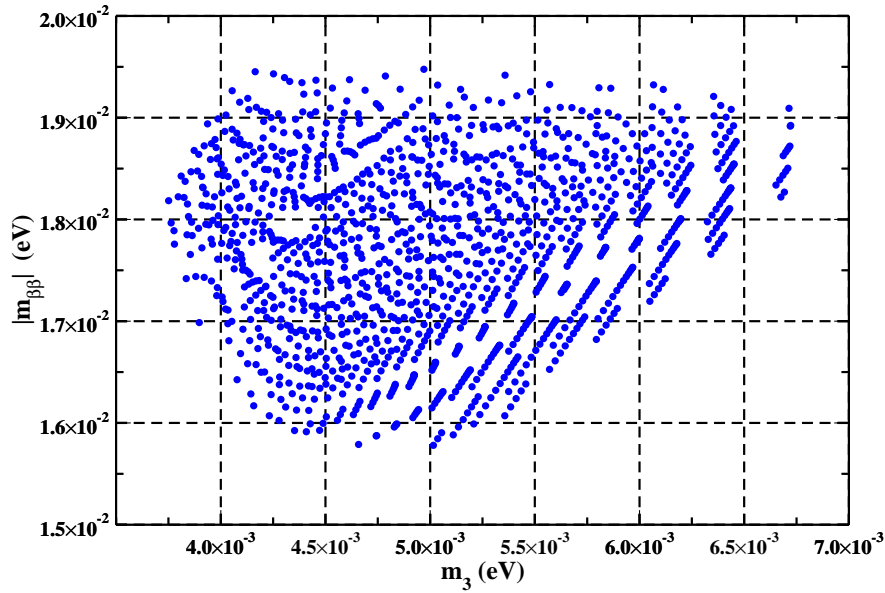}
		\caption{Points in the $m_{\beta\beta} - m_3$ plane corresponding to the consistent range of mixing angles in the model.}
\label{mbb-m3-dots}
\end{figure}

\begin{figure}
	\centering
		\includegraphics[angle=-90,width=0.8\textwidth]{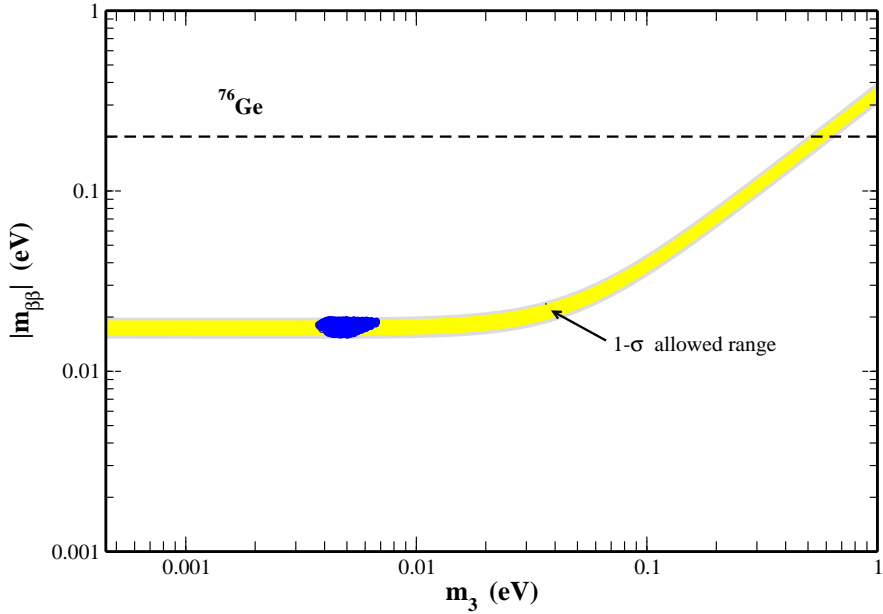}
		\caption{Allowed range for  $|m_{\beta\beta}|$ from oscillation data (region between gray lines), with the small spot being the points in Figure \ref{mbb-m3-dots}. The dashed line represents the smallest (most optimistic) upper limit from the absence of a direct observation of $0\nu\beta\beta$ decay in $^{76}$Ge ($|m_{\beta\beta} | < $(0.20-0.32) eV) \cite{Bilenky:2011tr}.}
\label{allowedrange}
		\end{figure}

In Figure \ref{mbb-m3-dots}, we show the values for $|m_{\beta\beta}|$ as a function of $m_3$, that we obtain using the angles shown in Figure \ref{lepangNeg}, and in Figure \ref{allowedrange} we show the $1-\sigma$ experimental allowed range for IH, together with the small region of points corresponding to our model. 

Note that the free parameter $y_l$ obtained in the charged lepton sector also plays a role in the neutrino sector. Before we found that in order to fit the charges lepton masses it was necessary for $y_l$ to lie within the range $0.0696 \leq |y_l| < 0.969$, but if in addition one requires $|m_{\beta\beta}|$ in the allowed experimental range, the region for $y_l$ gets reduced to $0.956 \leq |y_l| \leq 0.969$

It is worth mentioning that the $Z_4$ symmetry has also been implemented as flavor symmetry in reference \cite{EmmanuelCosta:2011jq}, in the framework of a SU(5) Grand Unified Theory, 
to obtain the NNI mass matrix form for quark and charged lepton sector. At low energy, below the GUT scale, the model reduces to the Two Higgs doublet model.
 In that extension, quarks charge assignment follows also reference \cite{Branco:2010tx}, and to conciliate with the SU(5) GUT group new conditions must be satisfied by the fermionic SU(5) multiplets, while the Higgs doublets $\Phi_1$  and $\Phi_2$ preserve the charges of their respective quintets. The lepton sector follows the assignment for down quark sector. Three right handed neutrinos fields are introduced as SU(5) singlets, with no constrained charges (they are free parameters), and the type-I seesaw mechanism is used to generate their masses. The effective neutrino mass matrix in this case, as in ours, does not exhibit the NNI form, and only two of six possible textures are found to fit well with experimental data. One of these two textures, named Texture-II$_{({12})}$ (IH) in \cite{EmmanuelCosta:2011jq},
has the same form as we found for neutrino mass matrix in expression (\ref{mnu}), but in our case with radiative mass generation and without the need of right handed neutrinos. This texture also demands the neutrino mass spectrum to have inverted hierarchy as is exhibited in our case.
Moreover, the $|m_{\beta\beta}|$ values we predict as a function of $m_3$, for the mixing angles we found, agree with their predictions showed in equation (\ref{su5}),  except that we have a bit lower $m_3$ values than their lower bound. 
\begin{equation}
\label{su5}
\begin{array}{c}
 \sin^2 \theta_{13} >0.010 , \\
0.0042 \text{ eV} \leq m_3 \leq 0.011 \text{ eV}, \\
0.015 \text{ eV} < |m_{\beta\beta}| < 0.022 \text{ eV}.\\
\end{array}
\end{equation}

\section{Conclusions}

We have taken the flavor structure of the quark sector presented in reference \cite{Branco:2010tx}, which requires the introduction of the $Z_4$ symmetry in the context of two Higgs doublet models, and extended it to construct a renormalizable flavor model for quark and lepton sectors.
Assuming that the charged leptons transform similarly as down quarks we obtain the NNI form for the mass matrix and we find the minimal requirements to generate neutrino masses radiatively. We verify that the theoretical values for masses and mixing angles provided by the model, are in agreement with the current experimental values, in particular with the last results of the T2K experiment \cite{Abe:2011sj,Oyama:2011pt}. The model exhibits IH for the neutrino mass spectrum and can fall within the allow region for $0\nu\beta\beta$ decay.


\begin{acknowledgments}
This work was supported in part by CONACYT and SNI (Mexico) and  A.~D.~R. acknowledges support by Fundaci\'on Carolina. 
 A.~D.~R. thanks the IFIC-C.S.I.C. (Universitat de Val{\`e}ncia) for its hospitality while part of this work was
carried out, to the IFIC-AHEP Group for giving some constructive suggestions, and also thanks Enrique D\'iaz for reading the manuscript.
\end{acknowledgments}


\end{document}